\documentclass[aps,pra,showpacs,eqsecnum,twocolumn]{revtex4}
\usepackage{graphicx}
\usepackage{amsmath,amssymb,amsthm,dsfont,bm,ulem}
\usepackage{color}

\begin{document}
\preprint{}
\title{Nonclassical detectors and nonclassicality}
\author{Alfredo Luis} 
\email{alluis@fis.ucm.es}
\homepage{http://www.ucm.es/info/gioq}
\author{Laura Ares}
\affiliation{Departamento de \'{O}ptica, Facultad de Ciencias
F\'{\i}sicas, Universidad Complutense, 28040 Madrid, Spain}
\date{\today}

\begin{abstract}
According to Born's rule quantum probabilities are given by the overlap between the system state and measurement 
states in a quite symmetrical way. This means that both contribute to any observed nonclassical effect that is
usually attributed just to the observed light state. This is relevant since typical measurement are highly non 
classical by themselves, such as  number states and quadrature eigenstates. We show that nonclassical effects only arise provided that the measurement is itself  nonclassical. Otherwise there is a classical-like 
model accounting for the observed statistics.
\end{abstract}
\pacs{42.50.Ar, 42.50.Xa,03.65.Ta }

\maketitle

\section{Introduction} 

Nonclassical effects are at the heart of the quantum theory. They are relevant both from fundamental as well as from practical 
reasons, since nonclassicality is actually a resource for future quantum technologies \cite{NCR}. 

Nonclassicality is  always revealed by  peculiar effects in observed statistics $p(m| \psi )$ where $\psi$ is  the system state and  $m$ the outcomes. According to Born's rule quantum statistics are determined in a  symmetrical way  by the system state $| \psi \rangle$ and the measurement states $| m \rangle$,  this is $P (m | \psi ) = | \langle m | \psi \rangle |^2$ \cite{G}, where typically  $| m \rangle$ are the eigenvectors of the measured observable. This raises the question of  whether the nonclassicality revealed by $P (m | \psi )$ is a property of the observable states  $| m \rangle$ or a property of the state  $| \psi \rangle$  being observed. This is specially pertinent since typically $| m \rangle$ are highly nonclassical states by themselves, say number states and infinitely squeezed states, in photon-number and quadrature 
measurements, respectively  \cite{OJ}. 

\bigskip

Despite this natural and simple remark, the non classicality has been always ascribed to the observed state.  A quite remarkable 
example is the photoelectric effect, that can be regarded as an observation of the light intensity. This is usually interpreted as a proof of the quantum nature of the light, although it  can be satisfactorily explained exclusively in terms of the quantum properties of the detector, this is to say that it admits a semiclassical explanation \cite{LS69,MSZ03,GZ05}.

\bigskip

In this work we demonstrate that the nonclassicality of the detector is a necessary condition to obtain nonclassical statistics $P (m | \psi )$. This can be particularized to some simple and common signatures of nonclassical light such as subPoissonian statistics, quadrature squeezing  and photon anti-correlations showing that they unavoidably requires detectors that are themselves nonclassical. 

\bigskip

Strictly speaking, a single-observable statistics $P(m | \rho )$ cannot reveal by itself nonclassical behaviour. This is because in classical physics we can always replicate any quantum probability distribution. The most clear quantum signature is the lack of a joint probability distribution for incompatible observables. This naturally includes as a particular case the flagship on nonclassical signatures is the lack of a {\it bona fide} Glauber-Sudarshan  $P (\alpha)$ distribution \cite{P}. This is not the only criterion. Actually this is a particular case of a more general approach which consider pathologies in the statistics of the joint measurement of any two observables \cite{AL16,LM17}, that can reveal nonclassical behavior even for states with {\it bona fide} Glauber-Sudarshan  $P (\alpha)$ distributions such as Glauber and SU(2) coherent states \cite{ncpn,qq,LM17,SL18}.

In this work we examine the role of the nonclassicality of the measurement  on the nonclassicality of observed statistics $P(m | \rho )$.

\bigskip

\section{Joint measurements} 

For completeness let us outline here the basic methods revealing non classicality in a joint 
measurement scenario \cite{AL16,LM17}. 

\bigskip

\subsection{Probability distributions}

In the most general case, joint measurements take place in an enlarged space with auxiliary degrees of freedom in a fixed and known state. The statistics can be properly represented in the system space by a positive operator-valued measure (POVM) $\tilde{\Delta} (x,y)$
\begin{equation}
\label{tPtr}
\tilde{p} (x,y ) =  \mathrm{tr} \left [ \rho \tilde{\Delta} (x,y)  \right ] ,
\end{equation}
and $\rho$ is the density matrix of the system state, $x,y$ are the outcomes in the measurement of two observables. More specifically we asume that the corresponding marginal POVMs $\tilde{\Delta}_X (x)$ and $\tilde{\Delta}_Y (y)$ 
\begin{equation}
\tilde{\Delta}_X (x) = \int dy  \tilde{\Delta} (x,y) , \quad \tilde{\Delta}_Y (y) = \int dx  \tilde{\Delta} (x,y) , 
\end{equation}
provide complete information about $X$ and $Y$, respectively. This is to say that 
there are functions $\mu_A (a, a^\prime )$ such that
\begin{equation}
\label{inv}
\Delta_A (a)  = \int d a^\prime \mu_A (a, a^\prime ) \tilde{\Delta}_A (a^\prime) ,
\end{equation}
$A=X,Y, \;\;\;  a=x,y$, 
where the functions $\mu_A (a, a^\prime )$ are state-independent and completely known as far as we know the measurement 
being performed, and $\Delta_A (a)$ are the exact, true POVMs corresponding to the system observables $A=X,Y$. 

\bigskip

The key idea is to extend this inversion (\ref{inv}) from the marginals to the complete joint distribution, to obtain a operator-valued measure \cite{AL16,LM17,WMM}:
\begin{equation}
\label{rD}
\Delta (x,y)  = \int d x^\prime d y^\prime \mu_X (x, x^\prime )   \mu_Y (y, y^\prime ) \tilde{\Delta} (x^\prime, y^\prime) .
\end{equation}
By construction the proper marginals are recovered 
\begin{equation}
\Delta_X (x) = \int dy  \Delta (x,y) , \quad \Delta_Y (y) = \int dx \Delta (x,y) .
\end{equation}

Likewise, all these relations hold between the corresponding probability distributions, in particular for the observed marginals $\tilde{p}_A (a)$, the inversion procedure 
\begin{equation}
\label{ptp}
\tilde{p}_A (a) =   \mathrm{tr} \left [ \rho \tilde{\Delta}_A (a)  \right ] , \quad p_A (a)  = \int d a^\prime \mu_A (a, a^\prime ) \tilde{p}_A (a^\prime) ,
\end{equation}
and, finally, for the inferred joint distribution 
\begin{equation}
\label{Ptr}
p (x,y ) =  \mathrm{tr} \left [ \rho \Delta (x,y)  \right ] ,
\end{equation}
so that 
\begin{equation}
\label{rP}
p (x,y)  = \int d x^\prime d y^\prime \mu_X (x, x^\prime )   \mu_Y (y, y^\prime ) \tilde{p} (x^\prime, y^\prime) .
\end{equation}
This is where the difference between classical and quantum physics emerges. As recalled below, in classical physics this program derives always in a {\it bona fide} joint probability distribution $p (x,y) $. This is not the case in quantum physics, where 
$p (x,y)$ may not exist or take negative values as a clear signature of nonclassical behaviour. 

This procedure includes all the other known approaches to nonclassical light such as pathological Glauber-Sudarshan distribution $P(\alpha)$. This is because $P(\alpha)$ pretends to be a joint distribution for incompatible field quadratures and is always determined via some kind of inversion procedure that provides always a {\it bona fide} probability distribution in the classical regime. 

\bigskip

\subsection{Classical physics}

Let us show that the inferred distribution obtained from the  inversion procedure (\ref{rP})  
leads always to a \textit{bona fide} probability distribution $p (x,y) $. 

\bigskip

Classically, the state of the system can be completely described by a legitimate probability distribution  $p ( \bm{\alpha})$, 
where $\bm{\alpha}$ are  all admissible states for the system, i. e.,  the corresponding phase space.

So the observed joint statistics can be always expressed as
\begin{equation}
\label{jj}
\tilde{p} (x,y) = \int d^2 \bm{\alpha} \tilde{X} (x | \bm{\alpha})\;  \tilde{Y} (y | \bm{\alpha})  p ( \bm{\alpha}) ,
\end{equation}
where  $\tilde{A}(a | \bm{\alpha}) $ is the conditional probability that the observable $\tilde{A}$ 
takes the value $a$ when the system state is $\bm{\alpha}$. By definition, phase-space points 
$\bm{\alpha}$ have definite, noncontextual  values for every observable so the factorized product of 
conditional probabilities $\tilde{X} (x | \bm{\alpha})  \tilde{Y} (y | \bm{\alpha})$ holds. Strictly speaking 
they are the product  of delta functions. In any case, this means that all $\tilde{A}(a | \bm{\alpha}) $ exist, 
are nonnegative, and no more singular than a delta function. So Eq. (\ref{jj}) expresses the separability of 
any joint measurement in classical optics. 

By construction, we know that there are $\mu_A (a,a^\prime )$ functions so that the analog of  Eq. (\ref{ptp}) 
holds and we get the exact conditional probabilities 
\begin{equation}
\label{rel}
 A( a | \bm{\alpha}) = \int d a^\prime \,\mu_A (a,a^\prime )\;  \tilde{A} ( a^\prime | \bm{\alpha}) .
\end{equation}
where  $A (a | \bm{\alpha}) $ is the conditional probability that the observable $A$ 
takes the value $a$ when the system state is $\bm{\alpha}$. 

Thus, because of the separable form for the observed joint statistics in Eq.  (\ref{jj}) we readily get from 
Eqs. (\ref{rP}) and (\ref{rel}) that the result of the inversion is  the actual joint distribution for $X$ and $Y$
\begin{equation}
\label{legi}
p (x,y)  =  \int d^2 \bm{\alpha} \, X(x | \bm{\alpha})\;  Y (y | \bm{\alpha}) p ( \bm{\alpha}).
\end{equation}

\bigskip

Thus, lack of positivity or any other pathology of the retrieved joint distribution $p (x,y) $ is then a signature 
of nonclassical behaviour. In turn, a necessary condition for this pathological behaviour of $p (x,y) $ is the 
lack of separability of the observed joint probability distribution $\tilde{p} (x,y ) $.

\bigskip

\section{There is no nonclassical light without nonclassical detectors} 

Let us express the inferred POVM (\ref{rD}) in the Glauber-Sudarshan representation as
\begin{equation}
\label{DA}
\Delta  (x,y) = \int d^2 \alpha p (x,y | \alpha )  | \alpha \rangle \langle \alpha | , 
\end{equation}
where $| \alpha \rangle$ are the Glauber coherent states. Then 
\begin{equation}
p (x,y ) =  \pi \int d^2 \alpha \; p (x,y | \alpha )   Q_\rho (\alpha ) ,
\end{equation}
being $Q_\rho (\alpha )$ the Husimi Q function of $\rho$ 
\begin{equation}
\label{Qrho}
Q_\rho (\alpha ) = \frac{1}{\pi} \langle \alpha| \rho | \alpha \rangle .
\end{equation}

We are attempting to describe classical measurements within a quantum scenario. The most natural way is to mimic the classical structure of statistics in Eq. (\ref{legi}), this is to say that $p (x,y | \alpha )$ factorizes with {\it bona fide} conditional probabilities 
\begin{equation}
p (x,y | \alpha )=p_X ( x | \alpha ) p_Y (y | \alpha ) , \quad p_A ( a | \alpha ) \geq 0 ,
\end{equation}
and $Q_\rho (\alpha )$ playing the role of the classical $p ( \bm{\alpha})$. 
These are actually the conditions considered when setting a classical-like scenario to derive the Bell inequalities. 

\bigskip

With this the joint distribution becomes  
\begin{equation}
\label{ccm}
p (x,y ) =  \pi \int d^2 \alpha p_X ( x | \alpha ) p_Y (y | \alpha )  Q_\rho (\alpha ) .
\end{equation} 
A key point here is that for every $\rho$ the function $Q_\rho (\alpha )$ exists and is nonnegative $Q_\rho (\alpha ) \geq 0$.  Thus, for classical detectors we have always a classical-like hidden-variable model where $Q_\rho (\alpha )$ plays de role of a joint distribution over the classical variables $\alpha$ so that $p (x,y ) $ is always a {\it bona fide} probability distribution an there is no nonclassicality. This is to say, that there are nonclassical effects only if the observables measured are themselves nonclassical, this is that $p_A (a |\alpha)$ do not exist, take negative values or are more singular than delta functions. In other words, nonclassical distributions $p(x,y)$ are obtained only if  the POVM elements $\Delta  (x,y)$ are more nonclassical than Glauber coherent states. 

\bigskip

The other way round we may consider the Glauber-Sudarshan representation for the system state $\rho$ exchanging the roles of state and observables in Eqs. (\ref{DA}) and (\ref{Qrho})  
\begin{equation}
\rho= \int d^2 \alpha  P_\rho \left ( \alpha \right )  | \alpha \rangle \langle  \alpha | ,
\end{equation}
so that the inferred joint distribution can be expressed as 
\begin{equation}
\label{cm}
p (x,y ) =  \pi \int d^2 \alpha P_\rho ( \alpha ) Q_{\Delta (x,y)}  (\alpha ) ,
\end{equation}
being \cite{LS99,RL09}
\begin{equation}
Q_{\Delta (x,y)}  = \frac{1}{\pi} \langle \alpha |   \Delta ( x,y)  | \alpha \rangle .
\end{equation}

Let us note that now there is an important difference with respect to Eq. (\ref{ccm}). This is that  $Q_{\Delta (x,y)}  (\alpha ) $ may not exist, be more singular than a delta function or take negative values. An example of each one of these possibilities can be found in Ref. \cite{LM17}: lack of existence for a suitably designed double homodyne detector, and negative for one-half spin-like measurements. Thus, there is room to observe nonclassical phenomena for states with well defined nonnegative $P_\rho (\alpha)$, such as coherent states. We may say that they are revealing the nonclassicality of the observation procedure, but let us recall that we have just demonstrated that nonclassical measurements is always  a necessary condition even if the system state is more nonclassical than coherent states. 

\bigskip

\section{Some simple examples} 

The above analysis shows that all quantum signatures must vanish whenever the measurement is classical and $\Delta (x,y)$ admits a well-behaved Glauber-Sudarshan distribution irrespective of the quantumness of the observed light state. Let us illustrate this point with some paradigmatic examples of nonclassicality.

\bigskip

\subsection{SubPoissonian statistics} 

A classic test of nonclassicality is subPossonian statistics, $\Delta^2 n < \langle  n \rangle$ which is incompatible with a {\it bona fide} $P_\rho (\alpha )$. Fur projection on number states the corresponding $p_N (n | \alpha )$ is extremely singular and thus nonclassical \cite{SU63}, this is 
\begin{equation}
| n \rangle \langle n | = \int d^2 \alpha p_N (n | \alpha ) | \alpha  \rangle \langle \alpha  |  .
\end{equation}
where $p_N (n | \alpha )$ is actually the Glauber-Sudarshan $P$ function of a number state being extremly singular containing derivatives of the delta function
\begin{equation}
p_N (n | \alpha ) = \frac{n! e^{|\alpha |^2}}{2 \pi | \alpha | (2 n)!} \frac{\partial^{2n}}{\partial | \alpha |^{2n}} \delta \left ( | \alpha | \right )  .
\end{equation}
Let us examine whether the subPoissonian behavior still holds  when we replace the above highly nonclassical $p_N (n | \alpha )$ by its classical version
\begin{equation}
\label{PN}
p_N \left ( n | \alpha \right ) = \delta  \left ( n - | \alpha |^2 \right ) ,
\end{equation}
where $n$ is now a continuous variable. In such a case the corresponding statistics reads
\begin{equation}
\label{pQ}
p_N (n | \rho ) = \frac{1}{2} \int_{2 \pi} d \phi Q_\rho \left ( \alpha = \sqrt{n} e^{i \phi} \right ) ,
\end{equation}
where we have used that $d^2 \alpha = (1/2) d |\alpha |^2 d\phi$ and $\phi$ is the phase of $\alpha$. 
This statistics can be related with the actual exact photon-number distribution $p_m$ in the form
\begin{equation}
\label{pc}
p_N (n) = \sum_{m=0}^\infty \frac{n^m e^{-n}}{m!} p_m,
\end{equation}
where $p_m = \langle m | \rho | m \rangle$ and $| m \rangle$ are number states, $\hat{n} | m \rangle = m | m \rangle$, being $\hat{n}$  
the number operator.

It is worth noting that this clearly resembles the formula for the number of photoelectrons recorded in a photoelectric scenario, just by replacing  the classical distribution for the integrated intensity by the discrete photon-number distribution $p_m$. The parallels are even more noticeable if we recall that  the photoelectron distribution arises in a semiclassical calculus, where the light is described classically while the detector is quantum. Here the situation is just the opposite, the detector is considered classical while the light is treated quantum-mechanically. Thus, this is some kind of {\it semiquantum} model. 

\bigskip

Taking into account that
\begin{equation}
\int_0^\infty  dn \; n^k e^{-n} = k! ,
\end{equation}
we readily get from Eqs. (\ref{pQ}) and (\ref{pc})
\begin{equation}
\langle n \rangle = \langle \hat{n}  \rangle + 1 ,
\end{equation}
and
\begin{equation}
\langle n^2 \rangle =   \langle \hat{n}^2  \rangle + 3 \langle \hat{n}  \rangle + 2,
\end{equation}
so that 
\begin{equation}
\Delta^2 n = \langle n^2 \rangle - \langle n \rangle^2 =  \Delta^2 \hat{n} + \langle n \rangle \geq \langle n \rangle .
\end{equation}
So, if we replace quantum by classical-like measurement all states are Poissonian (number states) or superPoissonian (all the rest).  
Therefore, we may safely say that subPoissonian statistics holds only if the measurement is subPoissonian itself,  and so non classical.

\bigskip

\subsection{Anticorrelation}
 
Next we  move from subPoissonian statistics to anticorrelation of photocounts in the typical scenario displayed in Fig. 1, as the 
flagship of quantum optics. There a single photon impinges on a lossless beam splitter and two joint intensity measurement are 
performed at the outputs of the beam splitter. Since the photon is indivisible, the detectors can never both  trigger simultaneously 
so that $\langle \hat{n}_1\hat{n}_2 \rangle =0$. This is maybe the most clear and simple evidence of the quantum nature of light 
\cite{GRA86}.

\begin{figure}
\begin{center}
\includegraphics[width=4cm]{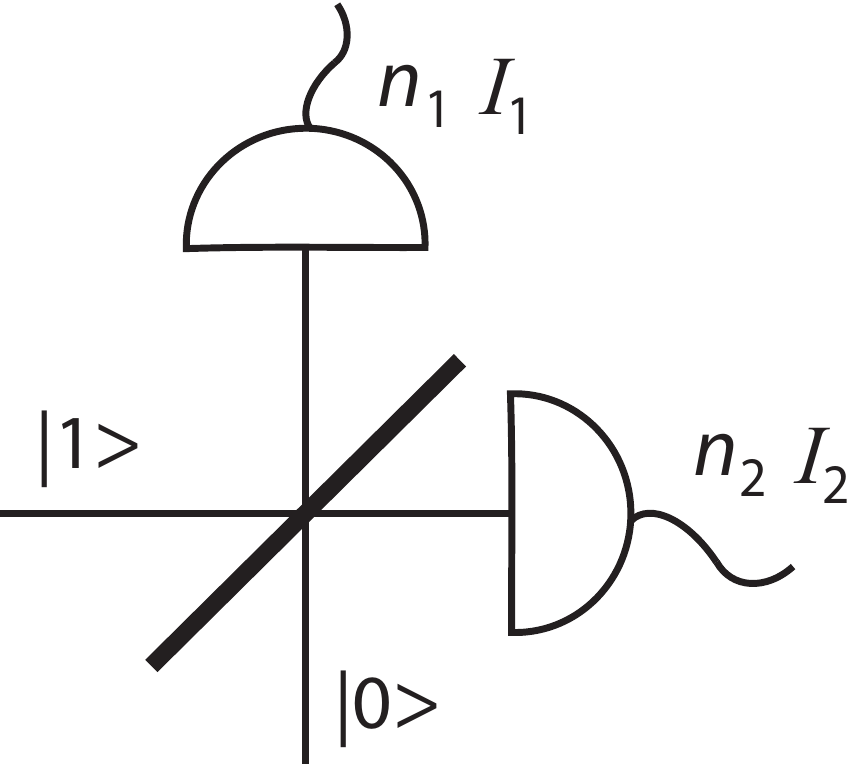}
\end{center}
\caption{Anti-correlation of photo counts for a single-photon input sate.}
\end{figure}

\bigskip

Thus we consider a two-mode version of the classical measurement described by the $P_N (n |\alpha )$ function in Eq. (\ref{PN})  to obtain the statistics when the field is in the one-photon state
\begin{equation}
p (n_1,n_2 | \rho ) = \left ( R n_1+ T n_2 \right ) e^{-n_1-n_2} ,
\end{equation}
where $R$, $T$ are the transmission and reflection coefficients with $T+R=1$. We simply get
\begin{equation}
\langle n_1 n_2 \rangle = 2 ,
\end{equation}
so that the alleged quantum effect would  be never observed if the detectors were classical-like devices.

\bigskip

Along the same lines we may examine the  Hong-Ou-Mandel effect illustrated in Fig. 2 \cite{HOM87}, where two photons impinge simultaneously on the input ports of a lossless 50 \% beam splitter. The quantum theory predicts  the result  $\langle \hat{n}_1 \hat{n}_2 \rangle =0$ again, as an evidence of the quantum nature of light.

\begin{figure}
\begin{center}
\includegraphics[width=4cm]{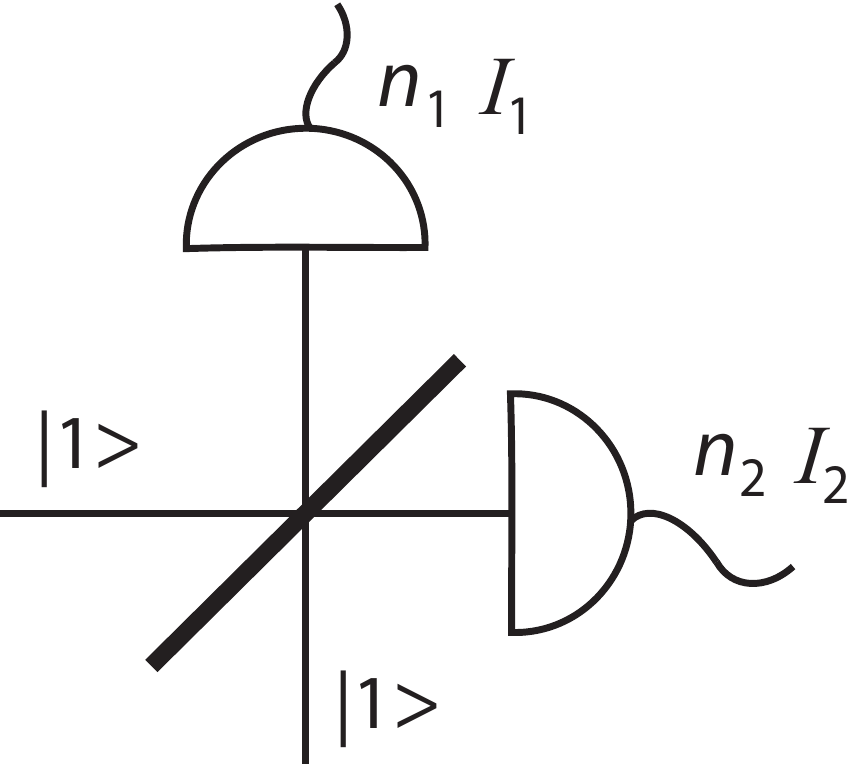}
\end{center}
\caption{Anti-correlation of photo-counts for a twin-photon input sate.}
\end{figure}

However, this result is not preserved if we replace the detectors by classical-like measurements as before, since 
the joint statistics would be:
\begin{equation}
p (n_1,n_2 | \rho) = \frac{1}{4} \left  (n^2_1+ n^2_2 \right )  e^{-n_1-n_2} ,
\end{equation}
leading to 
\begin{equation}
 \langle n_1 n_2 \rangle = 3 .
 \end{equation}

\bigskip

\subsection{Quadrature squeezing} 

As a further test of nonclassicality it is known that quadrature squeezing, $\Delta^2 x < 1/4$ where $x$ represents a field quadrature  $x = \Re \{ \alpha \}$, is incompatible with 
a {\it bona fide} $P_\rho (\alpha )$.  Here again the statistics of a quadrature measurement results by projection on the quadrature eigenstates, that are highly nonclassical being infinitely squeezed. We carry out the closest classical measurement by replacing the strongly quantum $p_X (x |\alpha)$ by
\begin{equation}
p_X \left ( x | \alpha \right ) = \delta  \left ( x -  \Re \{ \alpha \}  \right ) .
\end{equation}
so that the corresponding statistics reads
\begin{equation}
\label{px}
p_X (x | \rho ) = \int_{-\infty}^\infty dy  Q \left ( \alpha = x + i y \right ) ,
\end{equation} 
this is 
\begin{equation}
p_X (x | \rho ) = \sqrt{\frac{2}{\pi}} \int_{-\infty}^\infty d x^\prime e^{-2 (x-x^\prime)^2} p (x^\prime) ,
\end{equation} 
where $p(x)= \langle x |\rho  | x \rangle$, being $| x \rangle$ the quadrature eigenstates, is the true quantum quadrature distribution associated  to $\rho$.

Here again, computing the first two moments we get 
\begin{equation}
\langle x \rangle = \langle \hat{X} \rangle , \qquad  \langle x^2 \rangle = \langle \hat{X}^2 \rangle + \frac{1}{4} ,
\end{equation}
so that 
\begin{equation}
\Delta^2 x  = \Delta^2 \hat{X} + \frac{1}{4} \geq   \frac{1}{4} .
\end{equation}
Therefore, with classical quadrature measurements there would be no nonclassical behavior regarding this 
physical variable. 
 
\bigskip

\section{Conclusions} 
We have demonstrated that the nonclassicality of the detector is a necessary condition to obtain nonclassical statistics. We have shown explicitly that this is the case in the most typical signatures of nonclassical behavior, such as subPoissonian statistics, quadrature squeezing  and photon anti-correlations showing that they unavoidably requires detectors that are themselves nonclassical. Moreover we have demonstrated that it is possible to observe nonclassical phenomena for states with well defined nonnegative Glauber-Sudarshan distribution, such as coherent states. 

\bigskip

\section*{Acknowledgements}
L. A. and A. L. acknowledge financial support from Spanish Ministerio de Econom\'ia y Competitividad Project No. FIS2016-75199-P.  L. A. acknowledges financial support from European Social Fund and the Spanish Ministerio de Ciencia Innovaci\'{o}n y Universidades, Contract Grant No. BES-2017-081942.


\begin{thebibliography}{00}

\bibitem{NCR}
B. Yadin, F. C. Binder, J. Thompson, V. Narasimhachar, M. Gu, and M.?S. Kim, Operational Resource Theory of Continuous-Variable Nonclassicality, 
Phys. Rev. X {\bf 8}, 041038 (2018);
H. Kwon, K. Chuan Tan, T. Volkoff, and H. Jeong, Nonclassicality as a Quantifiable Resource for Quantum Metrology, Phys. Rev. Lett. {\bf 122}, 040503 (2019).

\bibitem{G}
In passing, this recalls the Goethe formulation of vision as the meeting of the inner and outer lights at the eye, as recalled by A. G. Zajonc, Goethe' s theory of color and scientific intuition, Am. J. Phys. {\bf 44}, 327--333 (1976).

\bibitem{OJ}
N. G. Walker,  Quantum Theory of Multiport Optical Homodyning, J. Mod. Opt. {\bf 34}, 15--60 (1987);
O. Jedrkiewicz, R. Loudon, and J. Jeffers, Retrodiction for coherent communication with homodyne or heterodyne detection, Eur. Phys. J. D {\bf 39}, 129--140 (2006).

\bibitem{LS69}
W. E. Lamb and M. O. Scully, Photoelectric effect without photons, discussing classical field falling on quantized atomic electron, in Polarization, Matter and Radiation,  Jubilee volume in honor of A. Kastler (Presses Universitaires de France, Paris, 1969).

\bibitem{MSZ03}
A. Muthukrishnan, M. O. Scully,  and M.  Zubairy, The concept of the photonÑrevisited, Optics and Photonics News TrendsÑThe Nature of Light: What Is a Photon? S-18--S-27  (2003) .

\bibitem{GZ05}
G. Greenstein and A. G. Zajonc, {\it The Quantum Challenge: Modern Research on the Foundations of Quantum Mechanics} (Jones \& Bartlett Learning, 2005).

\bibitem{P} 
L. Mandel and E. Wolf, \textit{Optical Coherence and  Quantum Optics}  (Cambridge University Press, 1995);
M. O. Scully and M. S. Zubairy, \textit{Quantum Optics} (Cambridge University Press, 1997).

\bibitem{ncpn}
J. Vaccaro, Number-phase Wigner function on Fock space, Phys. Rev. A {\bf 52},  3474--3488  (1995);
L. M. Johansen, Nonclassical properties of coherent states, Phys. Lett. A {\bf 329}, 184 (2004);
L. M. Johansen, Nonclassicality of thermal radiation, J. Opt. B: Quantum Semiclassical Opt. {\bf 6}, L21 (2004);
L. M. Johansen and A. Luis, Nonclassicality in weak measurements, Phys. Rev. A {\bf 70}, 052115 (2004);
A. Luis, Nonclassical polarization states, Phys. Rev. A {\bf 73},  063806 (2006).

\bibitem{qq}
A. Ferraro, L. Aolita, D. Cavalcanti, F. M. Cucchietti, and A. Ac\'{\i}n, Almost all quantum states have nonclassical correlations,
Phys. Rev. A {\bf 81}, 052318 (2010);
A. C. de la Torre, D. Goyeneche, and L. Leitao, Entanglement for all quantum states, Eur. J. Phys. {\bf 31}, 325--332 (2010).

\bibitem{AL16}
A. Luis, Nonclassical light revealed by the joint statistics of simultaneous measurements, Opt. Lett. {\bf 41}, 1789--1792 (2016).

\bibitem{LM17}
A. Luis and L. Monroy, Nonclassicality of coherent states: Entanglement of joint statistics, Phys. Rev A {\bf 96}, 063802 (2017).

\bibitem{SL18}
C. Sanchidri\'{a}n and A. Luis, Entanglement between total intensity and polarization for pairs of coherent states, Phys. Rev. A {\bf 97}, 043810 (2018).

\bibitem{LS99}
A. Luis and L. L. S\'{a}nchez-Soto, Complete Characterization of Arbitrary Quantum Measurement Processes, Phys. Rev. Lett. {\bf 83}, 3573--3576 (1999);
A. Luis, Quantum tomography of input-output processes, Phys. Rev. A {\bf 62}, 054302 (2000);
J. S. Lundeen, A. Feito, H. Coldenstrodt-Ronge, K. L. Pregnell, Ch. Silberhorn, T. C. Ralph, J. Eisert, M. B. Plenio, and I. A. Walmsley,
Tomography of quantum detectors, Nature Physics {\bf 5}, 27--30 (2009);
H. B. Coldenstrodt-Ronge, J. S.   Lundeen, K. L. Pregnell, A. Feito, B. J. Smith,  W. Mauerer, Ch. Silberhorn, J. Eisert, M.  B.  Plenio,  and  I. A.  Walmsley,
A proposed testbed for detector tomography, J. Mod. Opt. {\bf 56}, 432--441 (2009).

\bibitem{RL09}
A. Rivas and A. Luis, Nonclassicality of states and measurements by breaking classical bounds on statistics, Phys. Rev. A  \textbf{79}, 042105 (2009);
D. Mart'n and A. Luis, Nonclassicality in phase by breaking classical bounds on statistics, Phys. Rev. A {\bf 82}, 033829 (2010);
A. Luis, Nonclassicality tests by classical bounds on the statistics of multiple outcomes, Phys. Rev. A {\bf 82}, 024101 (2010);
A. Luis and A. Rivas, Independent nonclassical tests for states and measurements in the same experiment, Phys. Scr. \textbf{T143},  014015 (2011);
A. Luis, Nonclassicality in the statistics of noncommuting observables: Nonclassical states are more compatible than classical states,  Phys. Rev. A {\bf 84}, 012106 (2011);
A. Luis and A. Rivas, Angular-momentum nonclassicality by breaking classical bounds on statistics, Phys. Rev. A {\bf 84}, 042111 (2011).

\bibitem{WMM}
W. M. Muynck, \textit{Foundations of Quantum Mechanics, an Empiricist Approach}, 
(Kluwer Academic Publishers,  2002);
W. M. de Muynck and H. Martens, Neutron interferometry and the joint measurement of incompatible observables,
Phys. Rev. A \textbf{42}, 5079--5085 (1990);
W. M. de Muynck, Information in neutron interference experiments,
Phys. Lett. A \textbf{182}, 201--206 (1993); 
W. M. de Muynck,  An alternative to the L\"{u}uders generalization of the von Neumann projection, and its interpretation,
J. Phys. A: Math. Gen. \textbf{31}, 431--444 (1998).

\bibitem{Lee91}
C. T. Lee, Measure of the nonclassicality of nonclassical states, Phys. Rev. A {\bf 44}, R2775-R2778 (1991);
C. T. Lee, Theorem on nonclassical states, Phys. Rev. A {\bf 52}, 3374-3376 (1995);
N. L\"utkenhaus and S. M. Barnett, Nonclassical effects in phase space, Phys. Rev. A {\bf 51}, 3340-3342 (1995).

\bibitem{KB07}
J. Kofler and \v{C}. Brukner, Classical World Arising out of Quantum Physics under the Restriction of Coarse-Grained Measurements, 
Phys. Rev. Lett. {\bf 99}, 180403  (2007).

\bibitem{SU63}
E. C. G. Sudarshan, Equivalence of Semiclassical and Quantum Mechanical Descriptions of Statistical Light Beams, Phys. Rev. Lett. {\bf 10}, 277--279 (1963).

\bibitem{GRA86}
P. Grangier, G. Roger and A. Aspect,  Experimental Evidence for a Photon Anticorrelation Effect on a Beam Splitter: A New Light on Single-Photon Interferences,
Europhys. Lett. {\bf 1}, 173--179 (1986).

\bibitem{HOM87}
C. K. Hong, Z. Y. Ou and L. Mandel, Measurement of subpicosecond time intervals between two photons by interference, 
Phys. Rev. Lett. {\bf 59},   2044--2046 (1987).



\end{thebibliography}
\end{document}